\documentclass[12pt,preprint]{aastex}
\newcommand{\HI}{H~{\sc i}} 

\newcommand{\kms}{${\rm km~s^{-1}}$}

\shortauthors{MCCLURE-GRIFFITHS ET AL} 
\shorttitle{A DISTANT SPIRAL ARM IN THE MILKY WAY}

\slugcomment{\hspace{-1cm}{\em To appear in ApJ Letters}}
\begin{document} 

\title{A Distant Extended Spiral Arm in the Fourth Quadrant of the Milky Way}

\author{N.\ M.\ McClure-Griffiths,\altaffilmark{1} John M.\ Dickey,\altaffilmark{2} B.\ M.\ Gaensler,\altaffilmark{3} and A.\ J.\ Green\altaffilmark{4}}

\altaffiltext{1}{Australia Telescope National Facility, CSIRO, PO Box 76, Epping NSW 1710, Australia; naomi.mcclure-griffiths@csiro.au} 
\altaffiltext{2}{Department of Astronomy, University of Minnesota, 116 Church St SE, Minneapolis, MN 55455; john@astro.umn.edu}
\altaffiltext{3}{Harvard-Smithsonian Center for Astrophysics, 60 Garden Street MS-6, Cambridge, MA 02138; bgaensler@cfa.harvard.edu}
\altaffiltext{4}{School of Physics, Sydney University, NSW 2006, Australia; agreen@physics.usyd.edu.au}

\authoraddr{Address correspondence regarding this manuscript to: 
                N. M. McClure-Griffiths
                ATNF, CSIRO
                PO Box 76
		Epping NSW 1710
		Australia }

\begin{abstract}
Using data from the Southern Galactic Plane Survey we present a possible
distant spiral arm in the fourth quadrant of the Milky Way.  The very
distinct and cohesive feature can be traced for over 70\arcdeg\ as the most
extreme positive velocity feature in the longitude-velocity diagram.  The
feature is at a Galactic radius between 18 and 24 kpc and appears to be the
last major structure before the end of the \HI\ disk.  We compare the
feature with a Galactic spiral model and show that it is well reproduced by
a spiral arm of pitch angle, $i \sim 9\arcdeg$.  The arm is quite well
confined to the Galactic plane, dropping at most 1 kpc below the Galactic
equator.  Over most of its length the arm is 1 - 2 kpc thick.
\end{abstract}

\section{Introduction}
\label{sec:intro}
Historically, neutral hydrogen (\HI) in the disk of the fourth quadrant of
the Milky Way has been understudied.  The last fully sampled survey of this
region was made with the Parkes 18m Radiotelescope in the 1970's with an
angular resolution of 48\arcmin\ \citep{kerr86}.  Recently the International
Galactic Plane Survey \citep{taylor03,mcgriff01a} consortium has
undertaken a large-scale project to produce a fully sampled survey of the
Galactic plane with one arcminute resolution.  For the fourth quadrant, in
particular, the Southern Galactic Plane Survey \citep[SGPS;][]{mcgriff01a}
constitutes more than an order of magnitude improvement in angular resolution
over previous surveys.

In the Southern Milky Way little is known about the full extent of the \HI\
disk and the shape of the \HI\ density fall-off with Galactic radius.
\citet*{knapp78} made an ambitious attempt to measure the full extent of the
\HI\ disk.  The idea behind their work was that for very large radii the
radial velocity of the gas approaches the projection of the local standard
of rest (LSR) velocity, $\Theta_0$, onto the line of sight.  For an extended
disk of \HI\ there should be a pile-up of gas at $v_{LSR} = \Theta_0 |\sin
l|$.  Despite several very deep integrations, \citet{knapp78} did not detect
a pile-up.  Subsequent attempts, including a recent one with the Green Bank
Telescope, have also failed to detect any significant pile-up of gas at
extreme velocities \citep*{lockman02}.

The outer disks of large spiral galaxies occasionally show clear \HI\
spiral arms extending to galactic radii in excess of 30 kpc, while the
stellar arms stop at radii of $\sim 10$ kpc.  A classic example is
M83, where the \HI\ spiral arms extend three times as far as the
stellar arms \citep{tilanus93}.  Unfortunately, the positions of the
\HI\ spiral arms in the outer Milky Way are not well known.  Early
studies of Galactic spiral structure suggest possible arms out to
distances of 20 kpc or more.  In particular, the \citet{kerr69a}
latitude-velocity diagrams show a ridge of \HI\ in the fourth quadrant
at extreme positive velocities.  \citet{kerr69b} maps this feature as
a non-continuous spiral arm from $l=210\arcdeg$ to $l=290\arcdeg$.
\citet{davies72} also interpreted the low longitude ($l\lesssim
260\arcdeg$) end of this feature as an outer spiral arm, possibly
associated with high-velocity clouds.  Despite these early studies,
there has been very little interpretation of this ridge and no
discussion of it beyond $l=290\arcdeg$.

Here we present new data from the SGPS on this extreme positive
velocity ridge from $l=260\arcdeg$ to $l=330\arcdeg$ and discuss its
structure. We argue that it is a cohesive structure and interpret it
as a distant \HI\ spiral arm at the edge of the Milky Way.  Using a
simple Galactic spiral model, we show that the shape is well-fit by a
spiral arm with a pitch angle of $i \sim 9\arcdeg$.

\section{Observations}
\label{sec:obs}
The data presented here are part of the Southern Galactic Plane Survey, a
survey of the \HI\ spectral line and 21 cm continuum emission in the fourth
quadrant of the plane of the Milky Way.  The \HI\ data product is a
combination of high resolution, interferometric data obtained from the
Australia Telescope Compact Array (ATCA) and low resolution data from the
Parkes Radiotelescope,\footnote{The ATCA and the Parkes Radio Telescope are
part of the Australia Telescope which is funded by the Commonwealth of
Australia for operation as a National Facility managed by CSIRO.} covering
the region $253\arcdeg \leq l \leq 358\arcdeg$, $|b| \leq 1\fdg5$.  The
lower resolution Parkes survey covers an extended latitude range of $|b|\leq
10\arcdeg$.  A complete discussion of the observational and data reduction
techniques is given by \citet{mcgriffphd} and will be described further in
McClure-Griffiths et al.\ (2004, in prep.).  The final \HI\ cubes, made by
combining deconvolved images from Parkes and the ATCA in the Fourier domain,
have an angular resolution of $\sim$ 2 arcminutes, a channel spacing of
$0.82$~\kms\ and rms brightness temperature sensitivity per channel of $\sim
1.3$ K.
\section{Results}
\label{sec:results}
\begin{figure}[!ht]
\centering
\plotone{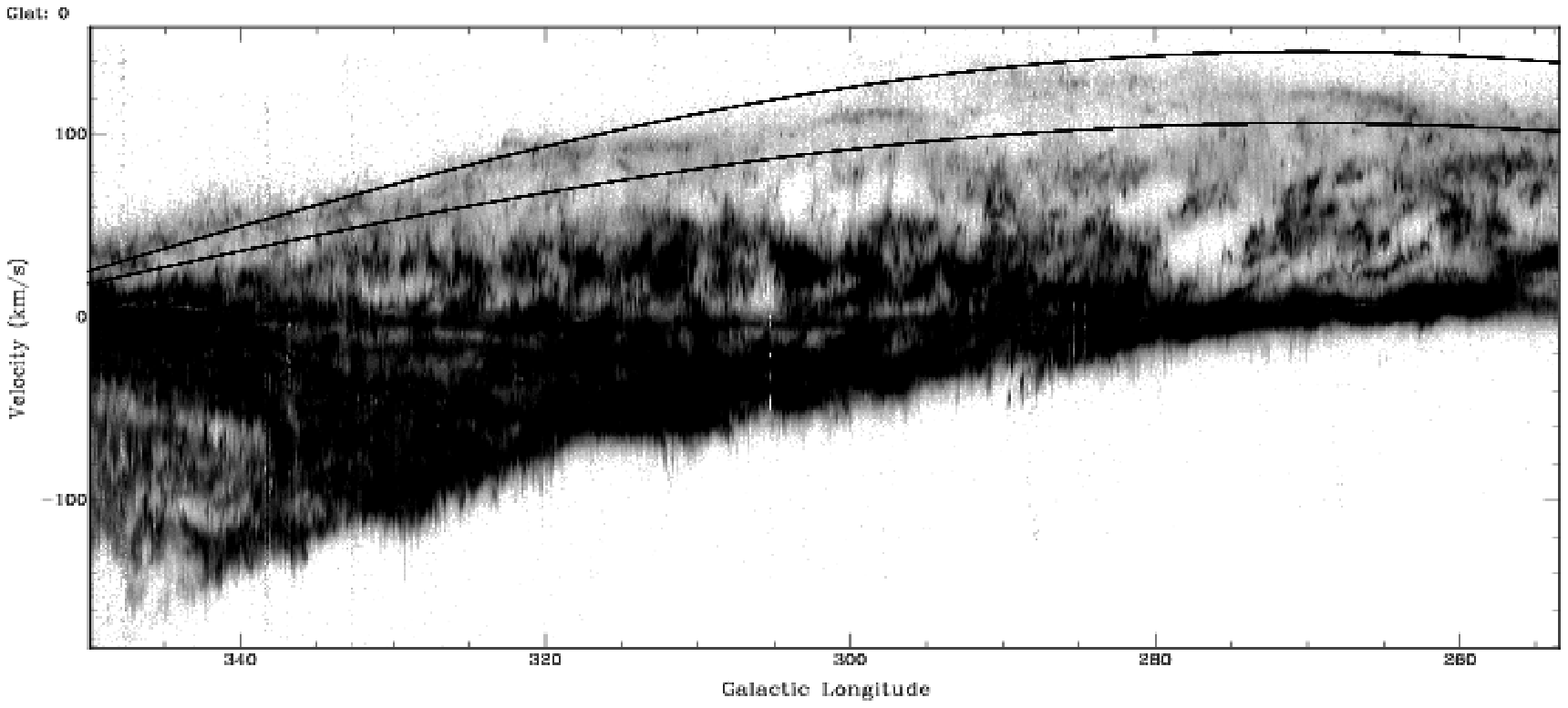}
\caption[]{\HI\ longitude-velocity ({\em l-v}) diagram of the Southern
Galactic Plane, showing the outer arm.  The arm is observed as a ridge
of emission at the most extreme positive velocities.  The image has an
angular resolution of 2 arcmin, a spectral resolution of $0.8$ \kms,
and uses a square-root transfer function from 1 K to 70 K.  The solid
lines are at constant galactocentric radii of 16 kpc and 24 kpc.
\label{fig:lv}}
\end{figure}

Combining data from all fields in the SGPS we created a composite
longitude-velocity ({\em l-v}) diagram at $b=0\arcdeg$ of the region
$253\arcdeg \leq l \leq 350\arcdeg$, shown in Fig.\ \ref{fig:lv}.  Examining
Fig.\ \ref{fig:lv}, we see that at the most extreme positive velocities there
is a thin ridge of emission that arcs from $l=253\arcdeg$, $v=102$ \kms\
through $l=299\arcdeg$, $v=110$ \kms, to $l=321\arcdeg$, $v=88$ \kms, and
possibly extending as far as $l=350\arcdeg$.  The feature is kinematically
distinct from the surrounding gas, rising to brightness temperatures of $T_b
\sim 15$ K over velocity widths on the order of $\Delta v \sim 9$ \kms.  It
is notably the last feature in the {\em l-v} diagram; beyond this the
profiles extend with long tails into the noise.  The feature is relatively
cohesive over the entire observed longitude range, with the only notable
exception between $l=275\arcdeg$ and $l=295\arcdeg$, where the arc shifts
approximately 10 \kms\ to higher velocities.

Overlaid on Fig.\ \ref{fig:lv} are lines of constant galactocentric
radius at $R_g = 16$ kpc and 24 kpc, assuming a \citet{brand93}
rotation curve with the IAU recommended values for the LSR, $\Theta_0=
220{\rm km s^{-1}}$, and Galactic center distance, $R_0= 8.5$ kpc. The
feature extends from about $R_g = 17$ kpc (heliocentric distance,
$d=16$ kpc) at the low longitude end to about $R_g =25$ kpc ($d = 26$
kpc) at the high longitude end.  At these radii the orbital period of
the feature must be on the order of 500 Myr.  If we assume that the
gas in the feature is optically thin, we can estimate a lower limit to
the average \HI\ number density along the midplane of the feature of
$n_{HI} \approx 4\times 10^{-2}~{\rm cm^{-3}}$.  Integrating over the
$z$-width of the feature, we find average \HI\ surface density of
$3\times 10^{-3}~{\rm M_{\sun}~pc^{-2}}$.  This is almost certainly an
underestimate of the total surface density.

Using Gaussian fits to the velocity profiles along the feature, we have
measured the velocity width in 48 directions where the emission is
strongest.  These fits cover the longitude range $l=280\arcdeg$ to
$l=325\arcdeg$, with a 10 degree gap around $l=293\arcdeg$, where the
feature is too weak to be reliably fit.  We find that the velocity width of
the feature has a mean of $9.6\pm 3.2$~\kms\ with no significant azimuthal
or radial variation.  The measured velocity width is slightly larger than
values obtained for the velocity dispersion in the outer Galaxy, which are
on the order of 5 \kms\ \citep[e.g.][]{blitz91}.  We therefore assume that
the velocity profiles are broadened by the spatial extent of the feature,
$dr$, times the velocity gradient, $dv/dr$.  If we assume that the velocity
dispersion is 5 \kms, then the upper limit of the physical line-of sight
width of the feature is $\sim 2$ kpc along most of its length.

\begin{figure}
\centering
\includegraphics[angle=90,width={\textwidth}]{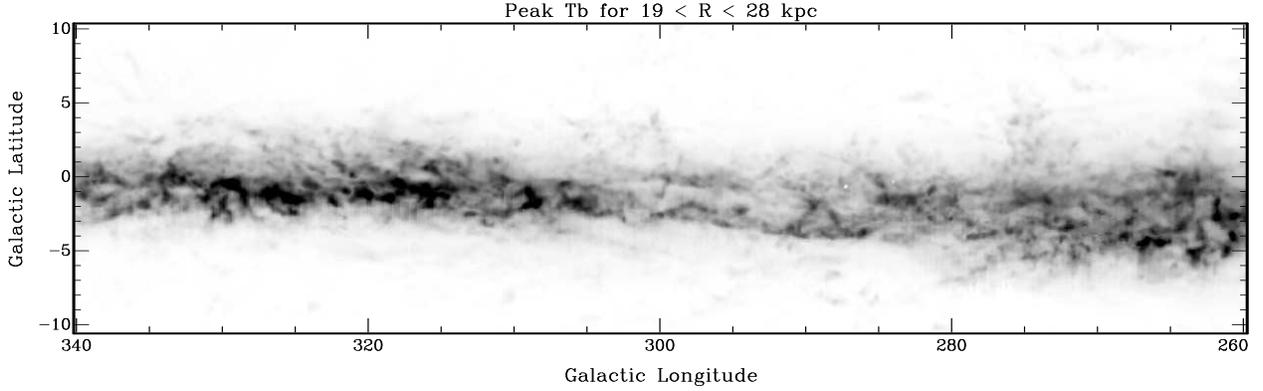}
\caption[]{Peak brightness temperature, $T_b$, as a function of
Galactic longitude and latitude for $19 \leq R_g \leq 28$ kpc.  The
grey-scale is linear from 1 K to 40 K.  The peak brightness
temperature traces the outer spiral feature, showing its latitudinal
variation with longitude.
\label{fig:peaktb}}
\end{figure}

The warp of the \HI\ disk in the fourth quadrant is such that at $R_g
\sim 20$ kpc the disk is coming back towards the Galactic midplane.
However, the feature is not restricted to $b=0\arcdeg$, but extends as
far as $\sim 5\arcdeg$ below the plane.  Fig.\ \ref{fig:peaktb} is a
grey-scale image of the peak brightness temperature observed with
Parkes (for latitude coverage $-10\arcdeg \leq b \leq +10\arcdeg$)
calculated for Galactic radii between $19 \leq R_g \leq 28$ kpc.  The
feature varies in latitude along its length, sweeping gradually from a
minimum central latitude of $b \approx -3\arcdeg$ ($z \approx -900$
pc) at $l=270\arcdeg$ to a central latitude of $b=0\arcdeg$ near
$l=320\arcdeg$.  At all longitudes there is significant emission at
$b=0\arcdeg$, which allowed us to detect the feature in our
high-resolution {\em l-v} diagram of $|b|\leq 1\fdg5$.  The full-width
at half-max in $z$ of the feature is relatively constant around
$\Delta z = 1.4$ kpc, varying between extremes of $\sim 1.2$ kpc and
$\sim 1.7$ kpc.  These values are consistent with, though somewhat
larger than, the \HI\ scale height at $2.1 R_0$ found by
\citet{merrifield92}.  Following \citet{merrifield92} we estimate that
a disk density of $\sim 6\times 10^{-3}~{\rm M_{\sun}~pc^{-3}}$ would
be required to confine the scale height of the feature. Though
\citet{merrifield92} notes that this density is consistent with the
e-folding length of the stellar disk, there is little evidence that
the stellar disk extends beyond 1.5 - 2$R_0$ \citep[e.g.][]{ruphy96}.
We therefore must assume that the required density for confinement is
provided by dark matter; an assumption supported by the dark matter
model of \citet{olling03}.

\section{Discussion}
\label{sec:discussion}
Here we consider several possible physical explanations for the
observed ridge.  A strong ridge of emission in the {\em l-v} diagram
is usually indicative of a density enhancement, a velocity
perturbation or a combination of the two.  Only for a very extended,
smooth \HI\ disk can a ridge of emission appear at the most extreme
velocity with no extra velocity perturbation.  In this case the final
feature of the {\em l-v} diagram is a pile-up in velocity space at the
projection of the LSR velocity onto the line of sight, following a
curve in {\em l-v} space of the form $v = \Theta_0 |\sin l|$
\citep{knapp78}.  We can easily exclude this explanation.  First, if
this effect were apparent one would expect to see a similar ridge in
the {\em l-v} diagram for the first and second Galactic quadrants,
which we do not.  Second, the velocities of the observed feature do
not agree with $v = \Theta_0 |\sin l|$. Assuming the IAU accepted
value for the LSR velocity of $\Theta_0 = 220$ \kms, we would expect
the ridge at $v=220$~\kms\ at $l = 270\arcdeg$ and at $v=190$ \kms\ at
$l=300\arcdeg$, whereas at those longitudes we observe this feature at
$v=123$ \kms\ and $v=108$ \kms, respectively.  To be consistent with a
disk pile-up in velocity space, the LSR velocity would need to be
reduced by almost a factor of two, to the unreasonable value of $\sim
125$ \kms.  The difference of Oort's constants $A - B$, which gives
the ratio $\Theta_0/R_0$, is well constrained by {\em Hipparcos}
measurements of Cepheid variables to be $\Theta_0/R_0 =
27.19\pm0.87~{\rm km~s^{-1}~kpc^{-1}}$ \citep{feast97}.  Reducing
$\Theta_0$ to $125$ \kms\ would force the Galactic center distance to
an unacceptable value of $4.5$ kpc.  It is therefore, highly unlikely
that this is the long sought pile-up of distant disk gas.

\begin{figure}
\centering
\plotone{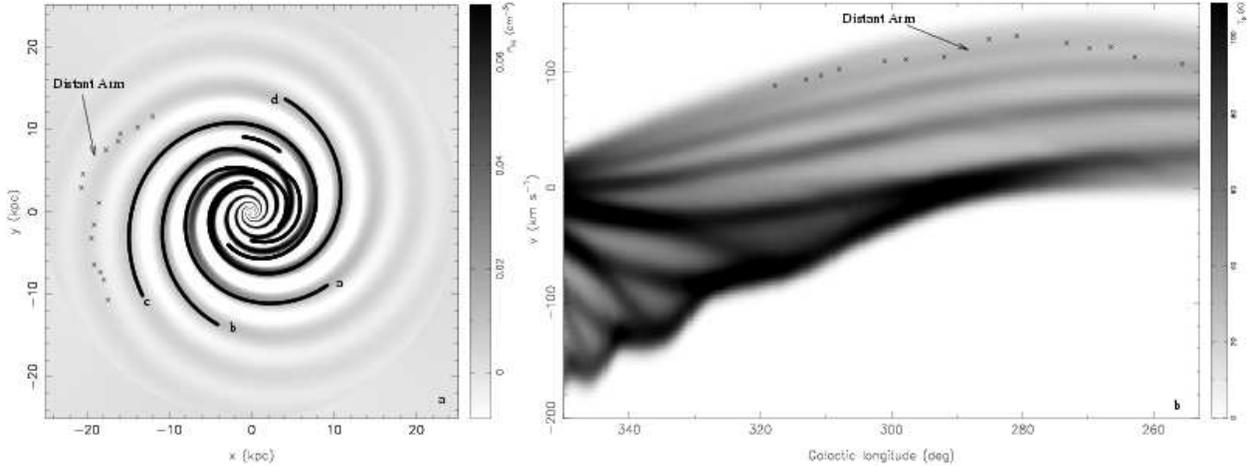}
\caption[]{Panel (a): Differential \HI\ density (spiral perturbation
minus the underlying Toomre disk) for the simple four-arm Milky Way
spiral model described in \S \ref{sec:discussion}.  The Sun is at
$(x,y)=(0~{\rm kpc},8.5~{\rm kpc}$). The crosses in both panels mark
observed positions of peak $T_b$ along the proposed spiral arm.  The
spiral model presented by \citet{cordes02} is overlaid as solid lines,
with the arms labelled as: {\bf a}: Scutum-Crux, {\bf b}:
Sagittarius-Carina, {\bf c}: Perseus, and {\bf d}: Outer.  Panel (b):
Synthetic {\em l-v} diagram created from the spiral pattern in panel
{\em b}.  The proposed distant arm in Panel {\em b} maps to a ridge of
emission at far positive velocities, as marked.
\label{fig:model}} 
\end{figure}

The most obvious cause of the feature is an outer spiral arm,
resulting in a slight (nominally a factor of two) density enhancement
and a significant ($\sim 10$ \kms) velocity perturbation, both of
which contribute to the creation of a ridge in the {\em l-v} diagram.
This hypothesis is supported by the feature's shape in {\em l-v}
space, which agrees well with a spiral arm found in a synthetic {\em
l-v} diagram of the Galactic spiral arms and by its relatively small
radial width ($< 2$ kpc), which suggests the action of a spiral
density wave that provides a radial confining force where there would
otherwise be none.  We created a simple synthetic {\em l-v} diagram,
following the recipe in \citet{burton71}, modified to assume a
four-arm spiral in agreement with the Galactic spiral models of
\citet{cordes02} and \citet{russeil03}.  We allow the pitch angle,
$i$, to vary such that $\tan i$ varies linearly from $i=16\arcdeg$ at
$R_g=4.25$ kpc to $i=10\arcdeg$ at $R_g = 18$ kpc.  The differential
\HI\ density of the model (spiral perturbation minus an underlying
Toomre disk) is shown in Fig.\ \ref{fig:model}({\em a}), overlaid with
the spiral model of \citet{cordes02}.  The synthetic {\em l-v}
diagram, shown in Fig.\ \ref{fig:model}({\em b}) uses this spiral
model, a \cite{brand93} rotation curve, and accounts for streaming
motions around the spiral arms.  The model is by no means definitive,
but it does reproduce many of the dominant features in the observed
{\em l-v} diagram.  In particular, the synthetic {\em l-v} diagram has
a narrow ridge of emission at the most extreme positive velocities.
The ridge corresponds to the spiral arm in the face-on diagram that
passes through $(x,y)=(-18~{\rm kpc},6~{\rm kpc})$.

Another possibility is that the shape of the feature in {\em l-v}
space can be explained by a non-circular motion of the LSR, as
suggested by \citet{blitz91}.  \citet{blitz91} estimate the magnitude
of the non-circular LSR to be $14 \cos l$ \kms.  Applying this
correction shifts outer Galaxy orbits away from the expected $\sin l$
shape.  With this correction the shape of the feature can be fit
reasonably well with an orbit of galactocentric radius 16 kpc.  Though
the shape of the feature can be fit by altering the LSR, the mere
existence of a ridge of emission cannot be explained by this motion.

A similar alternative explanation is that the outer Galaxy gas follows
elliptical orbits, as suggested by \citet{kuijken94}.  To estimate
this effect we included elliptical orbits in our model {\em l-v}
diagram.  \citet{kuijken94} found that Galactic \HI\ data are best fit
by an elliptical disk with the sun near the minor axis of the
potential and an equipotential axis ratio of 0.9.  Applying this
modification to our model {\em l-v} diagram results in an outer
envelope at positive velocities with a much steeper slope, $\Delta
v/\Delta l$, than is observed.  To fit the observed feature with an
elliptical orbit requires an axis ratio of 1.1, which
\citet{kuijken94} eliminate in their analysis.  Both \citet{kuijken94}
and \citet{blitz91} emphasize that there is little global evidence for
ellipticity at $R_g > 2 R_0$.  It therefore seems unlikely that the
feature can be explained by an elliptical potential.

A notable characteristic of this feature is that it shifts in velocity
by about 10~\kms\ between longitudes $l=275\arcdeg$ and
$l=295\arcdeg$, as seen in Fig.\ \ref{fig:lv}.  In the \citet{kerr69b}
sketches of this arm it is absent between these longitudes.
\citet{wannier72}, however, detect high latitude ($b\sim 10\arcdeg$)
emission at $v=153$ \kms\ at these longitudes and suggest that the
spiral arm arches to higher latitudes in this area.  Our data do not
show a shift towards higher latitudes, but rather a shift towards
higher velocities.  It is interesting that these longitudes agree well
with the Galactic longitudes of the Large and Small Magellanic Clouds
(LMC, SMC) and the Leading Arm of the Magellanic Stream.  In models of
the orbits of the Magellanic Clouds \citep[e.g.][]{gardiner96}, the
Clouds should cross the Galactic plane around $l=280\arcdeg$.  The
relative proximity of the LMC, in particular, should have a
significant dynamical effect on the outer Milky Way disk, and may
create a velocity perturbation in the outer spiral arm.  For example,
models of the effects of the Clouds on the Milky Way, such as that by
\citet{weinberg95}, show that the eccentric orbit of the LMC can
create radial velocity shifts on the order of 5 - 10 \kms\ beyond the
solar circle and in the direction of the Clouds.  The observed shift
in the outer arm velocity may be such an effect.

\citet{yanny03} recently detected a large diffuse, low latitude stellar
stream around the Milky Way between $122\arcdeg \leq l \leq 227\arcdeg$
around $R_g \sim 18$ kpc.  The nature of this feature is still uncertain
with suggestions that it is an outer spiral arm, a tidally disrupted
satellite galaxy or a resonance \citep{ibata03}.  \citet{frinchaboy04} find
several possibly associated stellar clusters that overlap the longitude
range of the \HI\ feature, but these have significantly smaller radial
velocities, are less confined to the Galactic midplane and are estimated to be
at $R_g \sim 10$ kpc, so we do not believe they are associated with the \HI\
feature.

\section{Conclusions} 
\label{sec:conclusions}
Using the SGPS we have mapped a large, coherent feature at extreme
positive velocities in the fourth quadrant of the Milky Way.  The
feature extends from $l=253\arcdeg$, $v=+102$ \kms\ to at least
$l=321\arcdeg$, $v=+88$ \kms, placing it at Galactic radii in excess
of 18 kpc.  The feature appears to be approximately 1 to 2 kpc thick
both in $z$ and along the line-of-sight.  We discuss several possible
explanations for the origin of the feature, specifically that it is a
pile-up in velocity space at $v = \Theta_0 |\sin l|$ due to an
extended, smooth \HI\ disk, and that it results from either a
non-circular LSR or an elliptical potential.  We exclude these and
conclude that the feature is best described by a very distant spiral
arm with a pitch angle of $i\sim 9\arcdeg$.

\acknowledgements This research was partially supported by NSF grant
AST-9732695 to the University of Minnesota.  We thank the referee for
raising several interesting points.

\clearpage

\end{document}